\documentclass[epj]{svjour}
\usepackage{graphicx}
\usepackage{amsmath}
\usepackage{amsfonts}
\usepackage{amssymb}
\usepackage{nicefrac}
\usepackage{subfigure}
\usepackage{color}
\usepackage{soul}

\renewcommand{\vec}[1]{{\mathbf{#1}}}

\graphicspath{{.}}

\begin{document}

\title{Convection in colloidal suspensions with particle-concentration-dependent
viscosity}
\author{Martin Gl\"assl \and Markus Hilt \and Walter Zimmermann\thanks{\email{walter.zimmermann@uni-bayreuth.de}}
}                     
\institute{Theoretische Physik I, Universit\"at Bayreuth, 95440 Bayreuth, GERMANY}

\date{submitted to European Physical Journal E (received 22 March 2010, received in final form 14 June 2010)}
%
\abstract{
The onset of thermal convection in a horizontal layer of a colloidal suspension is investigated in terms 
of a continuum model for binary-fluid mixtures where the viscosity depends on the local concentration of
colloidal particles. With an increasing difference between the viscosity at the warmer and the colder 
boundary the threshold of convection is reduced in the range of positive values of the separation ratio 
$\psi$ with the onset of stationary convection  as well as in the range of negative values of $\psi$ 
with an oscillatory Hopf bifurcation. 
Additionally the convection rolls are shifted downwards with respect to the center of the horizontal layer 
for stationary convection ($\psi>0$) and upwards for the Hopf bifurcation ($\psi<0$).
} 
\maketitle
%

\section{Introduction}\label{sec: intro}  

Rayleigh-B\'enard convection \cite{Benard:1900.1,LRayleigh:1916.2} is a prominent 
model for many processes in geoscience, atmospheric dynamics, and convection in technical
applications \cite{Cotton:2010,Houze:1994,BoPeAh:2000.1}. Moreover it is a classical
lab experiment for studying generic phenomena in nonlinear dynamics and pattern formation
\cite{BoPeAh:2000.1,CrossHo}. Typically the Oberbeck-Boussinesq (OB) approximation
is used for modeling thermal convection \cite{Legros:84,LandauVI,Busse:89.2,Gershuni:76}.
In this approximation  constant material parameters independent of the thermodynamic
variables are assumed, except for a temperature-dependent density in front
of the gravitational force, which is the essential driving force of convection.

An extension beyond the Boussinesq approximation is accounting for temperature-dependent viscosity
which is used, for instance, for modeling various phenomena in the Earth's mantle
\cite{Palm:1960.1,JensenO:1963.1,Turcotte:1970.1,Booker:1976.1,Booker:1982.1,Busse:85.1,White:1988.1,Christensen:1991.1,Yuen:1995.1,Tackley:1996.1}.
A spatially varying viscosity causes additional nonlinear couplings between
the temperature and the velocity field and breaks the up-down symmetry of the
flow field with respect to the center of the fluid layer. Often this symmetry breaking 
favors hexagonal patterns near the onset of convection.  A linear or sinusoidal
temperature dependence of the viscosity causes a reduction of the threshold
compared to the case of constant viscosity \cite{Palm:1960.1,JensenO:1963.1,Busse:85.1}.
For an exponential temperature dependence of the viscosity one can either have an
enhancement or a reduction of the onset of convection, depending on the viscosity contrast,
which measures the ratio of the viscosity at the boundaries \cite{Booker:1982.1}. Strongly
varying material properties in the Earth's mantle are a major motivation for using
a temperature-dependent viscosity in models of thermal convection in single component fluids
\cite{Christensen:1991.1,Yuen:1995.1,Tackley:1996.1,OgawaM:1991.1,OgawaM:2008.1}.
Materials in the interior of the Earth are however mixtures of several substances.
The dependence of the viscosity on the concentration of one component of the mixture
and its influence on the onset of convection is one possible effect to be investigated.

An interesting variant of the classical Rayleigh-B\'enard system represents
thermal convection in binary-fluid mixtures \cite{Legros:84}. Here the concentration
field of one of the two constituents enters the basic equations as an additional dynamic
quantity. Given the possibility of an oscillatory onset of convection, this enriches the
bifurcation scenario considerably and accordingly, the system has attracted wide attention 
by several reasons in the recent decades, including Non-Oberbeck-Boussinesq effects 
\cite{CrossHo,Linz:1987,Luecke:1998.1,Brand:84.1,Cross:88,Zimmermann:93.1}. 
In binary-fluid mixtures temperature gradients cause, via the Soret effect, concentration 
gradients and the concentration field couples via the buoyancy term into the Navier-Stokes
equations for the velocity field. In binary-fluid mixtures ({\it e.~g.} alcohol in water)
concentration diffusion is roughly a factor of hundred slower than thermal diffusion. 
For finite amplitudes of the velocity field, this leads to concentration variations in a narrow
boundary layer.

In recent experiments Rayleigh-B\'enard convection was investigated in colloidal suspensions
\cite{Cerbino:2002.1,Cerbino:2005.1,Cerbino:2009.1} with a special focus on Soret driven convection
\cite{Cerbino:2002.1,Cerbino:2005.1} and bistable heat transfer, involving new effects
like {\it e.~g.} sedimentation \cite{Cerbino:2009.1}. In theoretical studies of convection in colloidal suspensions
the model for binary-fluid convection is commonly used considering the colloidal nanoparticles as the
second fluid constituent \cite{Pleiner:2007.1}. Since nanoparticles are much larger than for instance alcohol 
molecules in water, their mass diffusion is more than two orders of magnitude smaller. Accordingly the 
Lewis number of suspensions reaches considerably smaller values, typically in the range of $10^{-4}$ \cite{Piazza:2008.2}.
The Soret coefficient, representing the strength of the Soret effect, grows linearly with the particle's size 
\cite{Piazza:2008.2,Braun:2006.1} and can therefore be changed in a wide range by varying mass density and size.

It is well known that the viscosity of a suspension rises with increasing concentration of
the solute particles \cite{LandauVI,Einstein:1906.1,Larson:99}.
The concentration of colloidal particles is sensitive
to temperature gradients via the Soret effect. 
Hence variations of the temperature cause spatial variations
of the particle concentration. This leads to a spatially varying viscosity
which is a function of the local concentration of the colloidal particles. 
The temperature dependence of the solvent's viscosity is neglected.
The goal of this work is the analysis of the onset of
convection  in a colloidal suspension in terms of 
a model for binary-fluid mixtures with a special emphasis 
on the effects of a particle-concentration-dependent viscosity.
In Sec.~\ref{model} we
briefly introduce the underlying equations of motion and the relation
between viscosity and particle concentration.
The linear stability analysis of the linear heat conducting state
 and the phase diagrams  for the onset of convection are presented in
in Sec.~\ref{linstab}. Our results are discussed and summarized in Sec.~\ref{conclusions}.

\section{Model}\label{model}  
Thermal convection in a horizontal layer of 
colloidal suspensions is described by the common mean field approach for binary-fluid mixtures 
\cite{Legros:84,Luecke:1998.1,Brand:84.1,Cross:88,Zimmermann:93.1,Guthkowicz:79}
with one extension: We take into account a linear dependence of the
viscosity on the local concentration of colloidal particles. 

Assuming that the mass density of the colloidal particles $\rho_c$
is similar to the mass density of the solvent $\rho_s$, {\it i.~e.} 
\begin{align}
\varepsilon = \frac{\rho_c}{\rho_s} \simeq 1\,,
\end{align}
the difference between $\rho_s$ and the mean density of the suspension $\rho_0$ is small
and therefore sedimentation effects are negligible.
In our case a spatially varying 
mass fraction of the colloidal particles $N(\vec{r},t)$ 
causes a spatially varying viscosity
similar to Einstein's law \cite{LandauVI}:
\begin{align}
\eta = \eta_0 \left(1+ \varepsilon \frac{5}{2} ~(N-N_0)\right)\,.\label{eq:Einstein} 
\end{align}
Effects of higher order in $ N(\vec{r},t)$ are neglected \cite{Larson:99,Batchelor:1972.1}.
Here $N_0$ represents the mean mass fraction of the colloidal particles and
$\eta_0$ describes the mean viscosity of the suspension.

The common set of basic transport equations for incompressible binary-fluid mixtures,
cf. Refs.~\cite{Legros:84,Luecke:1998.1,Brand:84.1,Cross:88,Zimmermann:93.1}, involves
the density of the mixture $\rho(\vec{r},t)$, the fluid velocity $\vec{v}(\vec{r},t)$, 
the temperature field $T(\vec{r},t)$, the mass fraction of the particles $N(\vec{r},t)$,
and the pressure field $p(\vec{r},t)$:
\begin{subequations}
\begin{align}
\nabla \cdot \vec{v} &= 0 \,, \label{eq:incompressible} \\
(\partial_t +  \vec{v} \cdot \nabla) \,T &= \chi \nabla^2 T\,,  \label{eq:heat} \\ 
(\partial_t +  \vec{v} \cdot \nabla) \,N &= D \left[\nabla^2  N + \frac{k_T}{T_0} \nabla^2  T \right]\,, \label{eq:continuity} \\
(\partial_t +  \vec{v} \cdot \nabla) \,\vec{v} &= -\frac{1}{\rho_0}\nabla  p + \nabla \cdot (\nu  \nabla)  \vec{v} 
+  \frac{ \rho}{\rho_0} \vec{g}\,. \label{eq:navier-stokes}
\end{align}
\end{subequations}
Eq.~\eqref{eq:incompressible} describes the incompressibility of the fluid. $\chi$ 
in the heat equation \eqref{eq:heat} denotes the thermal diffusivity of the mixture
and $D$ in the diffusion equation \eqref{eq:continuity} the diffusion constant. Due to the size of the colloidal particles $D$ takes much smaller
values than in molecular fluid mixtures. The dimensionless thermal-diffusion ratio $k_T$
representing the cross coupling between the temperature gradient and the particle flux
is related to the Soret coefficient $S_T$ via $k_T =  N(1- N) T S_T$ and can be either positive 
or negative. In the following $k_T \simeq N_0(1- N_0) T_0 S_T$
is regarded as constant. In our model the kinematic viscosity $\nu=\eta/\rho_0$ in
the Navier-Stokes-equations \eqref{eq:navier-stokes} is a function of space,
as introduced by Eq.~\eqref{eq:Einstein}.
The gravity field $\vec{g} = - g \vec{e}_z$ is chosen parallel to the z-direction.

For the local density $\rho$ of the suspension we use a linearized equation of state 
\cite{Legros:84,Gershuni:76},
\begin{align}
 \rho = \rho_0 \left[1-\alpha(T-T_0) + \beta(N-N_0) \right]\,,
\end{align}
with the thermal expansion  coefficient $\alpha = - (  1/\rho_0) \partial \rho /\partial T$ and
$ \beta = (1/\rho_0) \partial \rho/\partial N$ reflecting the density contrast between the solvent 
and the suspended particles. According to the Boussinesq approximation this dependency of the density is taken 
into account only within the buoyancy term.
The sign of $\beta$ indicates whether the colloidal particles have a higher or a lower mass density 
than the solvent. Here we assume  $\beta >0$, corresponding to $\varepsilon >1$.

{\it Boundary conditions.~}
The  fluid layer is confined between two impermeable and parallel plates at a distance $d$,
and extends infinitely in the $(x,y)$-plane. 
The lower plate ($z=-d/2$) is kept at a higher temperature $T_0 + \delta T/2$, the upper plate
($z=+d/2$) at a lower temperature $T_0 - \delta T/2$.

Considering realistic no-slip boundary conditions for the flow field the
boundary conditions at $z=\pm d/2$ are:
\begin{subequations}
\begin{align}
0 &=  v_x =  v_y=  v_z= \partial_z  v_z  \,, \\
0 &= \partial_z  N + \frac{k_T}{T_0} ~ \partial_z  T,  \\
 T&= T_0 \mp \frac{\delta T}{2}\,.
\end{align}
\end{subequations}
For geophysical applications free-slip boundary conditions 
\begin{align}
 0 &=  \partial_z v_x =  \partial_z v_y=  v_z= \partial_z^2  v_z  \,,
\end{align}
for the flow field are
often considered to be more realistic \cite{OgawaM:1991.1}. 
In the case of constant viscosity and free-slip boundary conditions
an analytical determination of the onset of convection is possible \cite{Brand:84.1}.
This advantage is lost when introducing a concentration dependent viscosity
and one has to rely on numerical methods.

{\it Basic state.~} 
In the absence of convection ($\vec{v}=0$) the
 time-independent heat-conducting state is described by:
\begin{subequations}
\label{heatcond}
\begin{align}
\label{eq:condT}
 T_{cond} &= T_0 - \delta T \,\frac{z}{d} \,,\\
\label{eq:condN}
 N_{cond} &= N_0 - \delta N \,\frac{z}{d} \quad \mbox{with} \quad
\delta N = - \frac{k_T}{T_0}~\delta T \,.
\end{align}
\end{subequations}
For further analysis it is convenient to separate the basic heat-conducting state in 
Eq.~(\ref{heatcond}) from convective 
contributions of the temperature and concentration fields as follows:
$ T(\vec{r},t) =  T_{cond}(z) +  T_1(\vec{r},t)$ and $ N(\vec{r},t) =  N_{cond}(z) +  N_1(\vec{r},t)$. 

Since we have rotational symmetry in the fluid layer 
we can restrict our further analysis concerning the onset 
of convection to two spatial dimensions,  namely to the $(x,z)$-plane.
 With this simplification the fluid velocity
can be expressed by a {\it stream function} 
$\Phi(x,z,t)$:
\begin{align}
 \tilde v_z = \partial_x \Phi ~,\quad \tilde v_x = -\partial_z \Phi\,.
\end{align}
Subsequently all lengths are scaled by the distance $d$ and
 time by the vertical thermal  diffusion time $d^2/\chi$.
Scaling the temperature field $T$ 
by $(\chi \nu_0) / (\alpha g d^3)$, the concentration field $N$ by $-(k_T \chi \nu_0) / (T_0 \alpha g d^3)$
and the stream function $\Phi$ by $\chi d$
we are left with {\it five dimensionless parameters}:
The {\it Rayleigh number} $R$, the {\it Prandtl number} $P$, the {\it Lewis number} $L$, and the
{\it separation ratio $\psi$},
\begin{align}
 P    = \frac{\nu_0}{\chi},\quad  
 L    = \frac{D}{\chi},\quad 
 R    = \frac{\alpha g d^3}{\chi \nu_0}\, \delta  T,\quad  
 \psi = \frac{\beta k_T}{\alpha T_0}\,,
\end{align}
are well known from molecular binary-fluid mixtures \cite{Cross:88,Zimmermann:93.1}.
The fifth dimensionless quantity
\begin{align}
 Q = \varepsilon~\frac{5}{2}\,\frac{\nu_0 \chi}{g \beta d^3}\,
\end{align}
is introduced to characterize the spatially varying
contribution to the viscosity of the suspension. For small particles in water
a reasonable value is $Q \simeq 0.01$
and for glycerin $Q\simeq 10$.
A further illustration of its physical meaning is obtained by considering the viscosity 
contrast $\tilde \eta = \eta(z=+\frac{1}{2})/\eta(z=-\frac{1}{2})$ between the 
concentration-dependent viscosity at the upper and the  lower boundary,
\begin{align}
 {\tilde \eta} &= \frac{1+\frac{5\epsilon}{4 T_0} \, k_T \delta T}
                       {1-\frac{5\epsilon}{4 T_0} \, k_T \delta T} =
\frac{1+\frac{1}{2} \, R \, \psi \, Q}{1-\frac{1}{2} \, R \,
\psi \, Q} \, ,
 \label{kontrast}
\end{align}
which is essentially a function of the product of
the three dimensionless control parameters: $R \, \psi \, Q$.
In the following we will discuss our results essentially in dependence of $\psi$ and $Q$
whereas $P$ and $L$ are regarded as constants.

For further analysis we introduce a rescaled temperature deviation 
$\theta = (R / \delta T) T_1$ and
a rescaled concentration deviation $\tilde N_1 =-(T_0 R / k_T \delta T)N_1$
in  terms of these dimensionless quantities.
In addition it is advantageous to introduce the combined function $\tilde c = \tilde N_1 - \theta$
instead of $\tilde N_1$.

Suppressing for reasons of simplicity all the tildes we obtain:
\begin{subequations}
\label{scaleeq}
\begin{align}
(\partial_t -\Delta) \theta - R \partial_x \Phi &= (\partial_z \Phi \partial_x - \partial_x \Phi \partial_z)\theta \,, \\
(\partial_t -L \Delta) c + \Delta \theta &=
(\partial_z \Phi \partial_x - \partial_x \Phi \partial_z)c \,, \\
( \partial_t - P\Delta) \Delta \Phi - P \psi \partial_x c &- P(1+\psi) 
\partial_x \theta \nonumber \\
\quad - \psi P Q R \left[z \Delta + 2\partial_z \right] \Delta \Phi &= (\partial_z \Phi \partial_x - \partial_x \Phi \partial_z) \Delta \Phi 
\nonumber \\
- \psi\,P Q ~[ \partial_x^2(c+\theta)\partial_x^2 \Phi
&+ 2 \partial_x \partial_z (c+\theta)\partial_x \partial_z \Phi  \nonumber \\
&+ \partial_z^2 (c+\theta)\partial_z^2 \Phi  ] \,.
\end{align}
\end{subequations}

No-slip, impermeable boundary conditions for the fields $\theta$, $c$, and $\Phi$ demand
\begin{align}
\label{bcsc}
 \theta = \partial_z c =\Phi=\partial_z \Phi=0 \quad \text{at $z= \pm\frac{1}{2}$}\, .
\end{align}
while free-slip, permeable boundary conditions \cite{Brand:84.1,Guthkowicz:79,Luecke:92}
\begin{align}
\label{bcsc2}
 \theta = c =\Phi=\partial_z^2 \Phi=0 \quad \text{at $z= \pm\frac{1}{2}$}\, .
\end{align}
All results presented in this work base upon calculations with no-slip, impermeable boundary conditions.
However, free-slip permeable boundary conditions lead to qualitatively comparable behavior.

\section{ \label{linstab} Linear stability of the heat conducting state and the
onset of convection.} 

The critical values of the control parameters at the onset of convection are 
determined by investigating 
the growth dynamics of small perturbations $\theta (x,z,t)$, $c(x,z,t)$,
and $\phi(x,z,t)$ with respect to the heat conducting state given
by Eqs.~(\ref{heatcond}).
The dynamics of such small perturbations is governed by the
linear parts of the coupled nonlinear equations (\ref{scaleeq}). They are solved by the
following exponential ansatz with respect to time $t$
and the horizontal $x$-direction:
\begin{align}
\left(\begin{array}{c}
  \theta(x,z,t)\\
  c(x,z,t)\\ 
  \phi(x,z,t) 
\end{array}
\right) 
&=\vec{u}_0(z) ~e^{iqx} e^{\sigma t}
\,, \\
\text{with}  \qquad \vec{u}_0(z) &=
\left(
\begin{array}{c} 
  \bar\theta(z) \\ 
  \bar c(z) \\
  \bar \phi(z)/(iq) 
\end{array} 
\right)\,.
\end{align}
With this ansatz and the eigenvalue $\sigma$ 
the linear parts of Eqs.~(\ref{scaleeq}) 
are transformed into
a boundary eigenvalue-problem with respect to $z$. 
The remaining linear ordinary differential equations (ODEs) are solved in this work by two different
methods.

The first one is a standard {\it shooting method} as described in detail for binary-fluid 
convection in   Ref.~\cite{Zimmermann:93.1}:
The resulting coupled ODEs for ${\bf u}_0(z)$ are integrated
by a standard solver for ODEs.
For every set of initial conditions a boundary determinant
$f(\sigma,R,q,Q,P,L,\psi)$ is calculated. 
During an iteration procedure $R$ or $\sigma$ are varied such 
that $f$ vanishes.
This yields $\sigma$ and $R$ as a function of the remaining parameters.

The second possibility is the so-called {\it Galerkin method} where the components
of ${\bf u}_0(z)$ are expanded with respect to a suitable chosen set
of functions which fulfill the boundary conditions, either Eq.~\eqref{bcsc} 
or Eq.~\eqref{bcsc2} 
- see, {\it e.~g.}, Ref.~\cite{Pesch:1996,Busse:1974,Canuto:1987}.
Here the linear ODEs are transformed into a generalized algebraic eigenvalue 
problem which can be solved numerically.
\begin{figure}[ttt]
  \begin{center}
  \subfigure[$\, \psi=10^{-4}$]
  {
  \includegraphics[width=0.9\columnwidth]{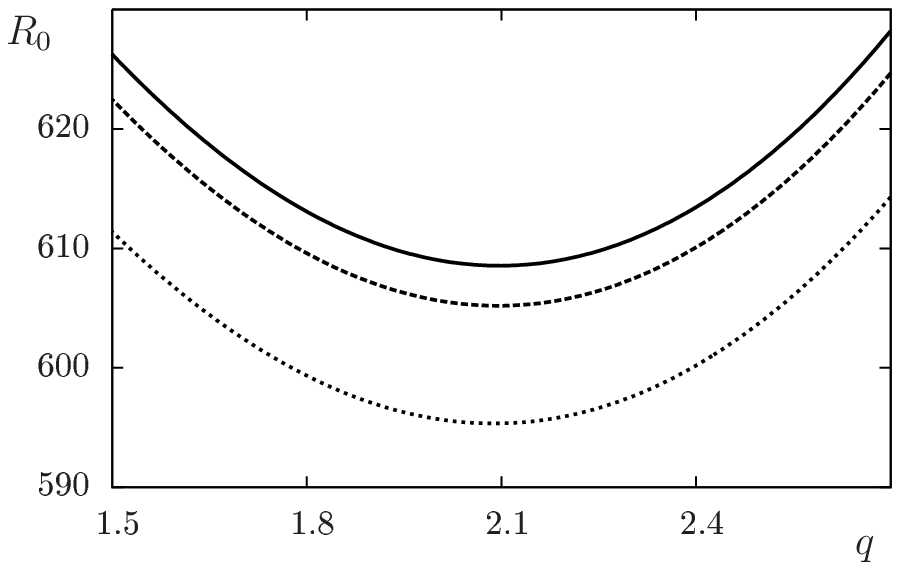}
  }
  \vspace{2mm}
  \subfigure[$\, \psi=10^{-2}$]
  {
  \includegraphics[width=0.9\columnwidth]{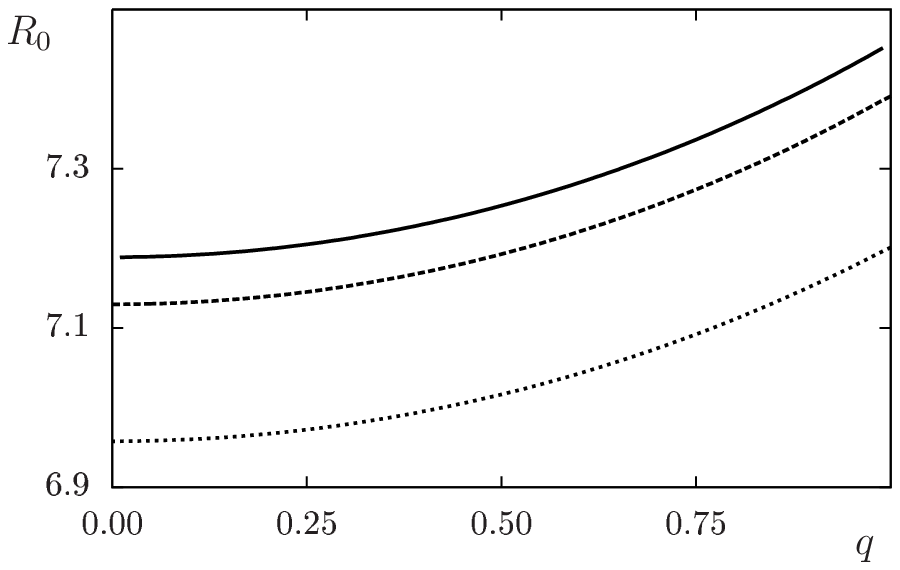}
  }
  \end{center}
\vspace{-5mm}
  \caption{Neutral curves $R_0(q)$ obtained according to Eq.~\eqref{neutcond}
 are shown for different values
of the viscosity parameter, $Q=0$ (solid line), $Q=5$ (dashed line) and  $Q=10$
(dotted line). In part a) for the positive
separation ratio   $\psi=10^{-4}$ and in b) for $\psi=10^{-2}$.}
\label{neutstationary1}
\end{figure}

At the onset of convection small 
perturbations $\theta$, $c$, and $\phi$ with respect to the
linear ground state neither grow nor decay, i. e. we are interested in the 
neutral stability condition for the fastest growing mode
(eigenvalue $\sigma$ with the largest real part),
\begin{align}
\label{neutcond}
\mbox{Re}(\sigma)=0\, \quad \text{with}\quad \sigma=\sigma(R,q,Q,P,L,\psi),
\end{align}
for the determination of the threshold.
Depending on parameters the system may show a stationary bifurcation with a vanishing
imaginary part of the eigenvalue, $\mbox{Im}(\sigma) = \omega=0$, or a Hopf bifurcation 
with a finite Hopf frequency $\mbox{Im}(\sigma) = \pm \omega \not =0$.

For a given set of material parameters the
neutral stability condition \eqref{neutcond} provides
the Rayleigh number $R_0(q)$ 
as a function of $q$, the so-called {\it neutral curve}
\begin{align}
\label{neutcurve}
 R_0(q)=R_0(q,Q,P,L,\psi)\,.
\end{align}
Similarly we obtain from Eq.~\eqref{neutcond} $\omega_0 = \omega_0(q,Q,P,L,\psi)$ 
in the case of a Hopf bifurcation.

In this work we choose a Prandtl number of $P=10$. 
Since the mass diffusion of the colloidal particles is much lower than
the thermal diffusion, a Lewis number of $L=10^{-4}$ is a reasonable
assumption. For small particles
in water $Q \simeq 0.01$ and in Glycerin $Q \simeq 10$ are appropriate
values. As the main changes for finite values of $Q$ occur if $\psi$
is comparable to $L$ we concentrate on rather small values of $\psi$.

Neutral curves for the present system are given in Fig.~\ref{neutstationary1} 
for two different positive values of the separation ratio $\psi$,
namely for $\psi=10^{-4}$ and $\psi=10^{-2}$, and for three different
values of the viscosity parameter $Q=0,5,10$.
The onset of convection is given by the minimum of the neutral curves $R_c = R_0(q_c) = \min R_0(q)$ 
which can be compared with its corresponding value in the case of a simple Newtonian 
fluid, $R_c^{N\!F} \simeq 1708$, and molecular binary mixtures \cite{Zimmermann:93.1}.

The reduction of the threshold in the range of $\psi > 0$ is well
known from molecular binary fluids \cite{Brand:84.1,Zimmermann:93.1} and
the reason is as follows.
Due to thermal expansion a fluid layer heated from below becomes lighter at the lower
boundary and there is a temperature induced density gradient. According to a positive 
Soret effect ($\psi > 0$) the colloidal particles tend to move to the upper plate 
which is colder. As the colloids are heavier than the molecules of the solvent
the Soret induced concentration gradient amplifies the temperature induced 
density gradient. Hence the fluid layer becomes unstable already at temperature 
differences much lower than for the onset of convection in a simple fluid.

For positive values of $\psi$, shown in Fig.~\ref{neutstationary1}, the bifurcation 
from the heat conducting state is stationary. While the minimum of the neutral curve
is located at a finite wavenumber $q_c > 0$ for values of $\psi$ that are small compared to $L$ (cf. 
Fig.~\ref{neutstationary1}(a)), the critical wavenumber
moves to $q_c = 0$ for larger values of $\psi$ (cf. Fig.~\ref{neutstationary1}(b)).
Finite values of $Q$ lead to a lowering of the neutral curves whereby the critical wavenumber
is nearly unaffected. The differences between the critical wavenumbers $q_c$ at the minimum 
of the neutral curves in Fig.~\ref{neutstationary1}a) is less than $0.1\%$.

\begin{figure}[ttt]
  \begin{center}
  \includegraphics[width=0.9\columnwidth]{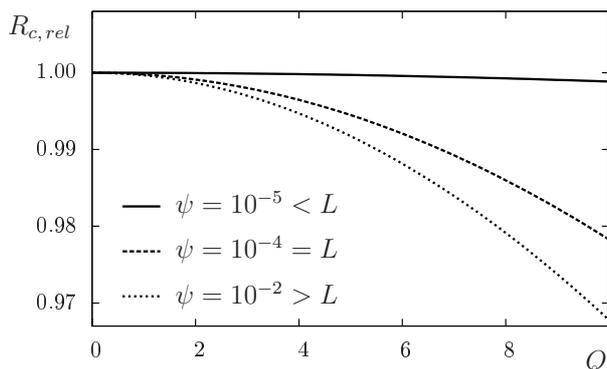}
  \end{center}
\vspace{-5mm}
  \caption{The relative critical Rayleigh number $R_{c,rel}= R_c(Q)/R_c(Q=0)$
is shown as a function of $Q$ for three different values of the separation ratio 
$\psi$: For $\psi=10^{-5} < L$ (solid) one has $q_c= 2.97$, for $\psi=10^{-4}=L$ (dashed) $q_c= 2.10$
and for $\psi = 10^{-2}>L$ (dotted) $q_c= 0.00$.
}
\label{neut_kf_Q}
\end{figure}

Fig.~\ref{neut_kf_Q} shows the threshold reduction as a function of $Q$ for three different 
positive values of $\psi$. Here we plot the critical Rayleigh number $R_c(Q)$ 
normalized by the critical Rayleigh number $R_c(Q=0)$ which we call the relative critical 
Rayleigh number 
\begin{align}
 R_{c,rel}(Q)= \frac{R_c(Q)}{R_c(Q=0)}\,.
\end{align}
The reduction of $R_{c,rel}(Q)$ as a function of $Q$ shows a parabola-like behavior as shown
in Fig.~\ref{neut_kf_Q}, which is accompanied by a tiny variation of the critical wavenumber 
$q_c$ along each curve. 
\begin{figure}[ttt]
  \begin{center}
 \subfigure[$\, \psi= 10^{-4}$] 
   {
  \includegraphics[width=0.45\columnwidth]{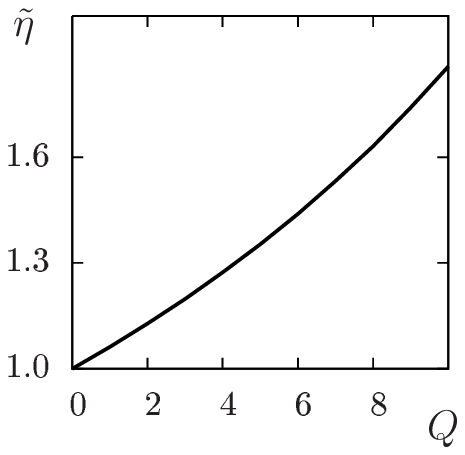}
   } 
 \subfigure[$\, \psi= - 10^{-4}$] 
   {
  \includegraphics[width=0.43\columnwidth]{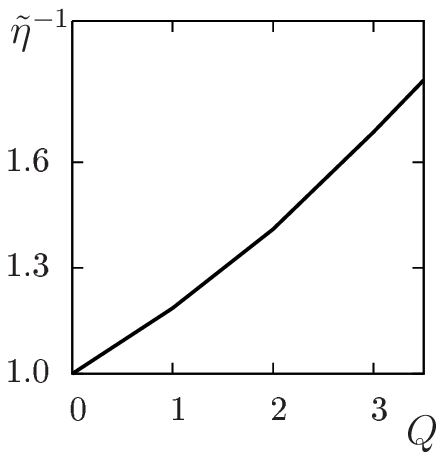}
   }
  \end{center}
\vspace{-5mm}
\caption{In part (a) the viscosity contrast $\tilde \eta$,
given by Eq.~\eqref{kontrast}, is plotted for $\psi=10^{-4}$ and in part b)
the inverse expression ${\tilde \eta}^{-1}$ is plotted for $\psi=-10^{-4}$. 
In both cases
the Rayleigh number $R$  entering Eq.~\eqref{kontrast} was taken
at the respective onset of convection.}
\label{Figcontrast}
\end{figure}

For finite values of $\psi$ the ground state displays 
a viscosity contrast $\tilde \eta \not =0$ as described by   Eq.~\eqref{kontrast}.
Fig.~\ref{Figcontrast} shows  $ \tilde \eta(Q)$
as a function of $Q$ for two different 
values of $\psi$ at the onset of convection $R=R_c$. 
The $Q$-range  in  Fig.~\ref{Figcontrast} was limited to the
range $|RQ\psi| \le 1/4$ where $k_T = const.$ is a reasonable assumption.

Another interesting question is how the threshold reduction behaves as function of the
viscosity contrast $\tilde \eta$.
Within our model the viscosity depends 
linearly on the concentration, according to Eq.~\eqref{eq:Einstein}. Since the concentration,
via Eq.~\eqref{eq:condN}, is
a linear function of the temperature, the viscosity is effectively
a linear function of the temperature too.
For a given viscosity contrast the threshold reduction of stationary convection ($\psi>0$) is
similar to that in a previous study of a single component fluid with a direct linear temperature dependence 
of the viscosity \cite{Busse:85.1}.
However, there are important differences between this single component study and the present model:
Here the viscosity contrast is caused by a well known physical mechanism, the Soret effect. 
The viscosity contrast is accompanied by heterogeneous distributions of the colloidal 
particles, which can be of relevance in the context of related convection phenomena in geoscience.
For a single component fluid in Ref.~\cite{Busse:85.1} one has only a stationary onset of 
convection. In contrast the concentration field for the colloidal particles represents an additional 
degree of freedom which, in the range of $\psi<0$, similar as in molecular binary-fluid 
mixtures \cite{Brand:84.1,Zimmermann:93.1}, leads to an oscillatory onset of convection via a 
Hopf bifurcation. 

Therefore arises a further interesting question: 
How does a finite viscosity contrast affect an oscillatory bifurcation?
We find in the range of $\psi<0$, as in the range of $\psi>0$, a lowering of the neutral
curve with increasing values of $Q$. The critical wavenumber $q_c$, as well as the Hopf frequency 
$\omega_c$, show only small and negligible changes when varying $Q$. 
The $Q$ dependence of the relative critical Rayleigh number $R_{c,rel}(Q)$ is slightly different from
its  behavior in the positive range of $\psi$. 
For $\psi>0$ the critical Rayleigh number $R_c(\psi)$ decays strongly as a function of $\psi$,
whereas the threshold increases only moderately for decreasing $\psi$ in the range $\psi<0$
\cite{Brand:84.1,Zimmermann:93.1}. Therefore, the viscosity contrast at threshold
given by Eq.~\eqref{kontrast} is for $\psi<0$ essentially a function of the product $\psi Q$.  
\begin{figure}[ttt]
  \begin{center}
  \includegraphics[width=0.9\columnwidth]{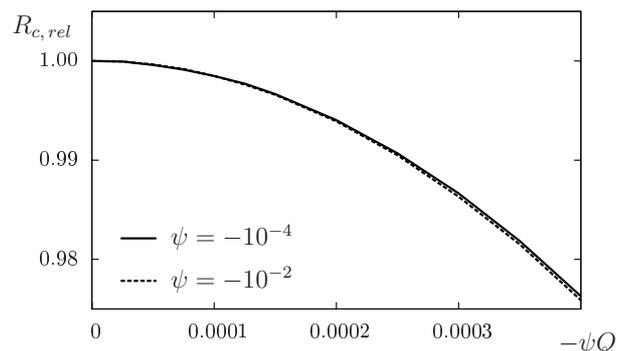}
  \end{center}
\vspace{-5mm}
  \caption{
The relative critical Rayleigh number $R_{c,rel}= R_c(Q)/R_c(Q=0)$
is shown as function of the product $- \psi Q$ for two different values
of the separation ratio $\psi$: For $\psi=-10^{-4}$ (solid)
and for $\psi=-10^{-2}$ (dashed). The critical
wavenumber along both curves takes the value $q_c \simeq 3.116$.
}
\label{neut_kf_QH}
\end{figure}

The relative threshold $R_{c,rel}(Q \psi)$ is shown in Fig.~\ref{neut_kf_QH}
as a function of the product $-Q\psi$ for two different values of $\psi<0$. The
two curves are nearly identical as expected.
Therefore, the changes of the threshold of the oscillatory convection are again 
essentially determined by the viscosity contrast $\tilde \eta$ given by
Eq.~\eqref{kontrast} in the restricted range $|R \psi Q| \le 1/4$.

\begin{figure}[ttt]
  \begin{center}
  \subfigure[$\, R_{c,rel}(\psi,Q)$]
  {
  \includegraphics[width=0.9\columnwidth]{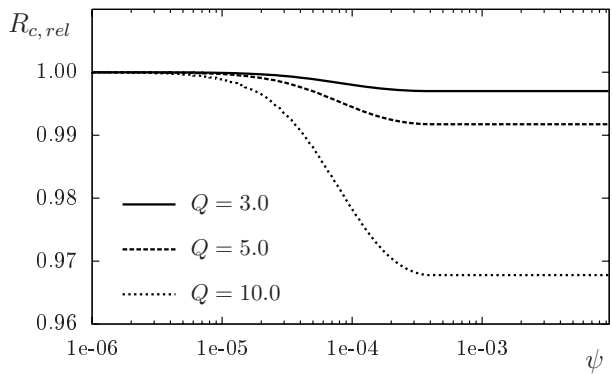}
  }
 \vspace{2mm}
  \subfigure[$\, q_c(\psi)$]
  {
  \includegraphics[width=0.9\columnwidth]{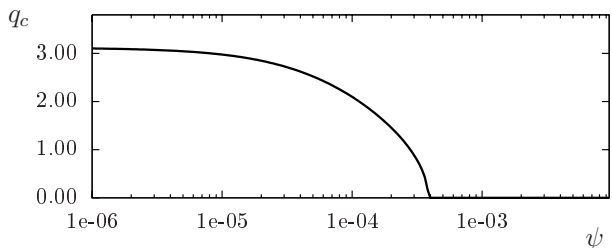}
  }
  \end{center}
\vspace{-5mm}
\caption{Part a) shows the relative
critical Rayleigh number $R_{c,rel}(\psi,Q)= R_c(\psi,Q) / R_c(\psi,Q=0)$ 
as a function of the separation ratio $\psi$ 
for three different values of Q:  $Q=3.0$ (solid), $Q=5.0$ (dashed),
and $Q=10.0$ (dotted). Part b) shows the corresponding critical wavenumber $q_c(\psi)$.
}
\label{neut_vers_psi}
\end{figure}

Another view on the parameter dependence of the threshold
is given in Fig.~\ref{neut_vers_psi},
where  the relative critical Rayleigh number $R_{c,rel}(\psi)$ is shown as
a function of the separation ratio $\psi$  
for three different values of the
viscosity parameter $Q=3.0, 5.0,$ and $10.0$. In the limit of
 a vanishing separation ratio $\psi$
the particle concentration and therefore the viscosity contrast
$\tilde \eta$ does not change by varying $Q$, so we have $R_{c,rel} \simeq 1$. 
For increasing values of $\psi$ the threshold drops down and finally reaches
a constant value, depending on $Q$. The strongest decay can be observed
in the region $\psi \simeq L$ accompanied by the critical wavenumber $q_c$
going down. $q_c$ finally vanishes at nearly the same value of 
$\psi \simeq 4 \times 10^{-4}$ independently of $Q$ as shown by Fig.~\ref{neut_vers_psi}b).

\begin{figure}[ttt]
  \begin{center}
  \subfigure[$\,$ velocity field]
  {
  \includegraphics[width=0.9\columnwidth]{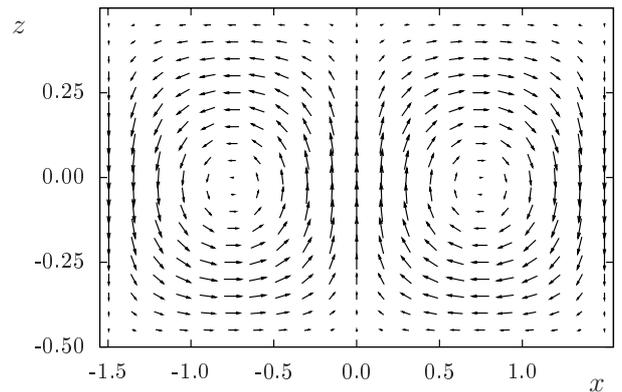}
  }
 \vspace{2mm}
  \subfigure[$\, v_x(z)$]
  {
    \hspace{0.15cm}
  \includegraphics[width=0.88\columnwidth]{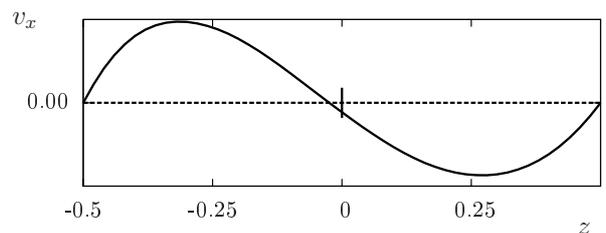}
  }
 \vspace{2mm}
  \subfigure[$\, z_0$ at $v_x(z_0)=0$]
  {
  \includegraphics[width=0.9\columnwidth]{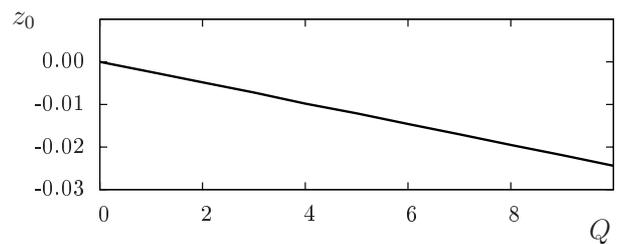}
  }
  \end{center}
\vspace{-5mm}
\caption{
The flow velocity field at the onset of stationary convection is 
shown in part a) for the parameters $\psi=10^{-4}$ and $Q=10$. 
Part b) depicts the variation of the velocity component $v_x(z)$ 
for the same parameters as a function of the vertical coordinate $z$
and part c) shows the shift of the position  $z_0$ where the
horizontal velocity vanishes, {\it i.~e.} $v_x(z_0)=0$, as a function of $Q$.
}
\label{flowprof}
\end{figure}

For $\psi>0$ an increasing viscosity contrast leads to a decrease of the velocity field 
at the upper boundary as the fluid motion tends to concentrate in the region
of lower viscosity at the lower boundary. Therefore the center of the convection rolls at $z_0$ is 
slightly shifted below the center of the convection cell at $z=0$ as shown 
in Fig.~\ref{flowprof}a) for $\psi=10^{-4}$. 
This shift $z_0$ is more clearly indicated in Fig.~\ref{flowprof}b), which shows 
the horizontal component of the convection velocity, $v_x(z)$, as function of $z$.
The shift $z_0$ to negative values increases with increasing values of $Q$ as shown 
in Fig.~\ref{flowprof}c).
This behavior is common for all positive values of $\psi$ with a finite value
of the critical wavenumber $q_c$. For $\psi<0$, in the range of an oscillatory
Hopf bifurcation, the viscosity is higher at the lower boundary and therefore 
the magnitude of the velocity is reduced near the lower plate.
Accordingly, the center of the convection rolls, $z_0$, is elevated with increasing
values of $\psi Q$ to $z_0>0$.
Our analysis shows that the shift $|z_0|$ depends essentially on the viscosity contrast 
$\tilde \eta$.

Assuming free-slip boundary conditions \eqref{bcsc2} instead of no-slip
boundary conditions \eqref{bcsc}, we obtain very similar results. For finite values
of $Q$ the threshold is reduced compared to its value at  $Q = 0$ and the center of the rolls
is shifted to the region of lower viscosity.

\section{Conclusions}\label{conclusions} 
The onset of thermal convection in a colloidal suspension 
was investigated in terms of a generalized continuum model for binary-fluid mixtures 
going beyond the common Boussinesq-approximation by
taking into account a linear dependency of the viscosity on the local 
concentration of the colloidal particles.

Spatial changes of the concentration of colloidal particles in the convection 
cell are induced by temperature variations via the Soret effect similar to that in 
molecular binary-fluid mixtures. The influence of the particle concentration on 
the viscosity is described by a dimensionless viscosity parameter $Q$,  
introduced in this work. 
For finite values of $Q$ the threshold is reduced compared to the limit $Q=0$,
for both the stationary bifurcation for $\psi>0$ and the oscillatory
Hopf bifurcation for $\psi<0$. 
This reduction depends essentially on the induced viscosity contrast $\tilde \eta$ 
at the choosen values of $Q$ and $\psi$.
We would like to emphasize that $|R \psi Q|$ may not be too large so that the 
linear approximation of the concentration variation in Eq. (7b) remains a 
reasonable assumption.

For both bifurcations the center of the convection rolls is 
shifted slightly away from the center of the convection cell: 
For $\psi>0$ the convection rolls are shifted below and for $\psi<0$  beyond the 
center of the cell, as the fluid motion always tends to concentrate in the region
of lower viscosity.
At the onset of convection the changes of the critical wavenumber of the convection 
rolls as a function of the dimensionless parameter $Q$ are rather small.

In a previous study on convection for a single component fluid in Ref.~\cite{Busse:85.1}
a linear dependence of the viscosity on the temperature was assumed
and its influence on the onset of stationary convection was investigated.
Assuming the same viscosity contrast a similar relative 
reduction in the threshold is obtained as in our model.
However, our model does not only describe a spatial variation of the material parameters. 
It describes also a mechanism leading, via a heterogeneous concentration distribution,
to a spatial dependence of the viscosity. This interrelation between concentration and 
viscosity variations may be of relevance to related phenomena in geophysical contexts. 
Additionally, the further degree of freedom in our model, introduced by the concentration 
dynamics, provides a richer bifurcation scenario and we can investigate the effects of 
a spatially varying viscosity on a stationary as well as on a Hopf bifurcation. 

In this work the viscosity $\hat \eta_0$ of the solvent of the colloidal
particles was assumed to be
constant. A natural extension of the present 
model would be to take also into account the effects of a temperature-dependent viscosity
of the carrier fluid. In the range of larger values of $\psi>0$ 
with small temperature differences at the threshold, this effect may be small.
However, in the range of small and negative values of $\psi$ this effect 
may play a role and one may then have a spatially  
varying mobility of the colloidal particles corresponding to a
temperature and spatially dependent Lewis number. 
A temperature dependent Lewis number is also expected for thermosensitive microgel
suspensions \cite{Ballauff:2007,Rehberg:2010} in which core-shell colloids
change their size as a function of the temperature and their mobility as well as 
the viscosity of the suspension.
This combined effect on the
viscosity  as well as the effects of nonlinear variations of the
particle concentration will be discussed elsewhere.

The vertical viscosity variation, caused by a temperature gradient and leading to
spatially dependent viscosity, 
breaks the up-down symmetry in the convection cell. 
Therefore the interesting question arises: In
which parameter range are hexagonal patterns favored 
for this binary-fluid model in the weakly nonlinear regime?

\begin{acknowledgement}
Stimulating discussions with F.~H. Busse, M. Evonuk, G. Freund, W. K\"ohler, W. Pesch, I. Rehberg,
and W. Sch\"opf are appreciated. This work has been supported by the German science 
foundation through the research unit FOR 608.
\end{acknowledgement}

\bibliographystyle{prsty}

\end{document}